\documentclass[10pt]{iopart}
\usepackage{iopams,bm,cite,color,graphicx,mathptmx,url}

\begin{document}

\title{Amplitude response of a Fabry-Perot interferometer} 

\author{Juan~J~Monz\'{o}n and Luis~L~S\'{a}nchez-Soto}

\address{Departamento de \'Optica, Facultad de F\'{\i}sica,
  Universidad Complutense, 28040~Madrid, Spain}

\date{\today}

\begin{abstract}
  The complex reflected and transmitted amplitudes from a Fabry-Perot
  interferometer are analyzed using a phase-space approach, in which
  the real and imaginary parts of those amplitudes are taken as basic
  variables. As functions of the optical path length of the cavity,
  the reflected amplitude describes a circle, whereas the transmitted
  is given by a hippopede. The system performance can be directly
  related to the geometrical parameters of these curves.
\end{abstract}

\eqnobysec


\section{Introduction}

The Fabry-Perot (FP) interferometer provides a superb illustration of
the mysterious ways in which interference works. Despite its apparent
simplicity, it plays a central role as high-resolution spectrometer,
laser resonator, or spectral filter, to cite only but a few of its
many relevant uses. A complete account of the subject can be found in
the two comprehensive monographs by Hernandez~\cite{Hernandez:1986zr}
and Vaughan~\cite{Vaughan:1989gf}.

This variety of fields of application spawned many descriptions of the
FP operation, each one capitalizing on specific aspects. The geometric
treatment, in which one adds the multiple beams reflected at each of
the different interfaces, is probably more instructive and,
accordingly, is reproduced in almost every
textbook~\cite{Born:1999yq}. The question can also be tackled by
imposing the appropriate boundary conditions, which gives the resonant
frequencies and allowed fields in the FP~\cite{Saleh:2007nr}. As the
boundary conditions appear as a linear system, they can pop up under
multiple guises: for example, the FP may be viewed as an optical
transmission system with feedback~\cite{Chen:2012hl}, or as a direct
application of the transfer
matrix~\cite{Yeh:2005eu,Sanchez-Soto:2012bh}, a method especially
germane to deal with layered structures.

Irrespective of the approach,  the focus is always on the
intensity distributions (both in reflection and transmission); namely, the
well-known Airy formulas. Even if the corresponding amplitudes are
somehow required to determine these distributions, they are ultimately
discarded  on the grounds that real experiments measure intensity.

This viewpoint can be challenged, however, by a discussion of the conventional
harmonic oscillator, wherein the complex amplitude is decomposed into
two orthogonal in-phase and out-of-phase quadratures, also known as
the dispersive and absorptive components~\cite{Crawford:1968aq}. These
quadratures are of paramount importance as they convey more useful
information than just the intensity. At the quantum level, for example, they play
the role of the effective position and momentum of the
oscillator~\cite{Schleich:2001qr}. Actually, a good deal of the
latest advances in quantum information processing stem from a proper
engineering of these quadratures, with homodyne detection constituting an
ideal tool for their measurement, whereas squeezing them provides an
efficient route to producing entanglement~\cite{Braunstein:2005cl}.

Inspired by this, we intend to shed light on the amplitude
response of the FP. Indeed, one can define an equivalent version of
the quadratures. When the parameters of the FP vary, the amplitudes
trace out elementary curves: the reflected amplitude is a circle,
and the transmitted one is a hippopede, a curve with remarkable
properties~\cite{Lawrence:1972db,Shikin:1995cr}. Furthermore, the FP
performance can be naturally assigned to the geometrical features of
these curves.

Apart from its elegance, this approach makes a close contact with
phase-space methods that pervade physics today.  The derivation is
straight and easy; suitable for undergraduates. Surprisingly enough,
to the best of our knowledge, such simple ideas have been not hitherto
explored; even if they can be important in some instances in which the
phase introduced by the FP matters,  as it happens in optical
metrology, where the stabilization is crucial.

\section{The Fabry-Perot: Basic background}

\begin{figure}[b]
 \centering{\includegraphics[height=5cm]{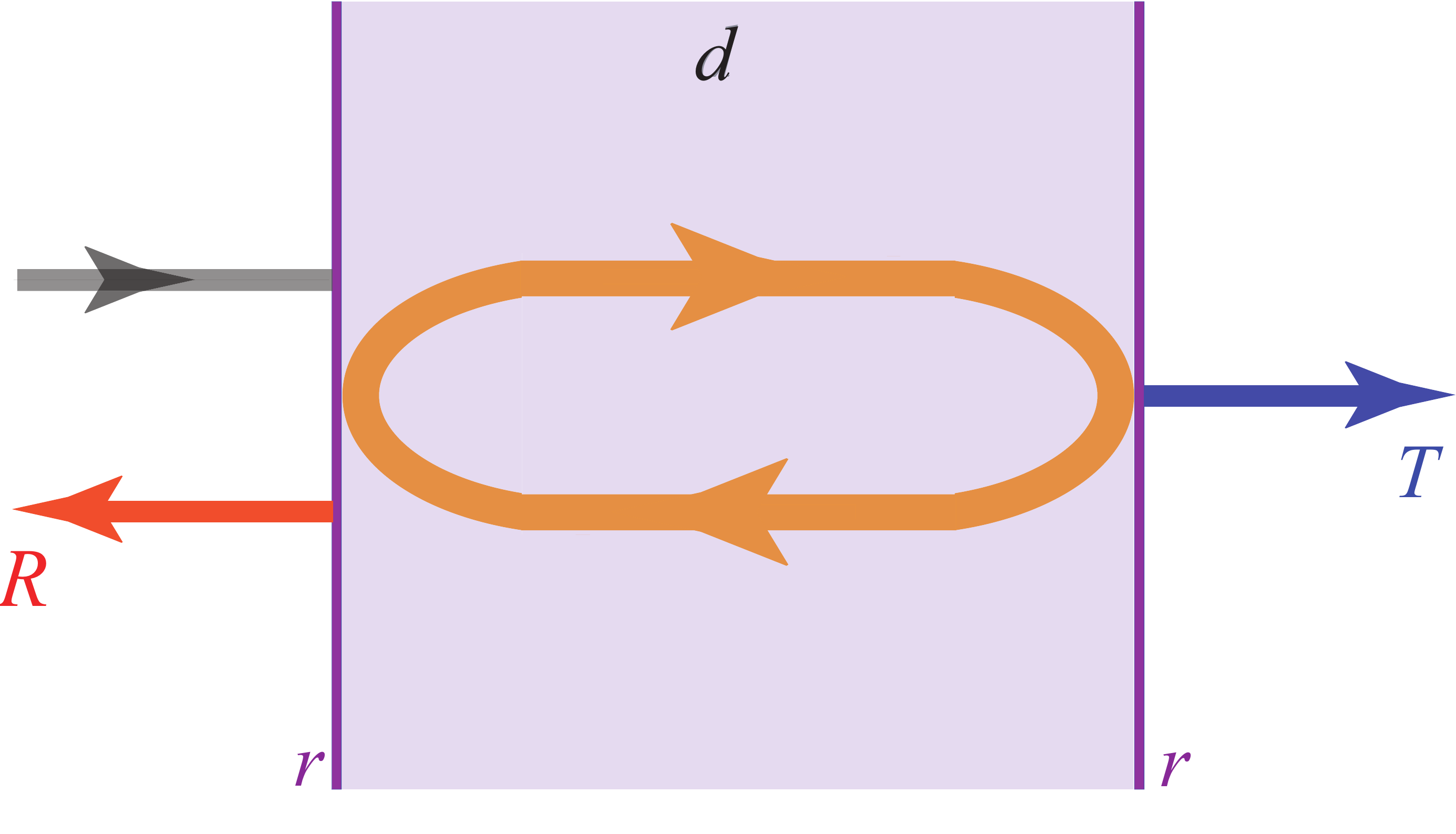}}
  \caption{Schematic block diagram of an FP. The intracavity fields
    can be analyzed from a variety of perspectives. The interferometer
   plate surfaces have reflection coefficient $r$ and  they are
   separated at a distance $d$. The reflected and transmitted amplitudes
   are labeled by $R$ and $T$, respectively, since we assume unit
   incident amplitude.}
  \label{fig:schema}
\end{figure}

The ideal FP interferometer consists of two parallel mirrors (that,
for simplicity, we assume to be identical) separated at a distance
$d$. Figure~\ref{fig:schema} is a block diagram of the system. This
can be addressed by considering a plane parallel plate of thickness
$d$ and refractive index $n$ immersed in a medium of index
$n^{\prime}$.  The plate is illuminated near normal incidence with a
linearly polarized quasi-monochromatic plane wave, with the electric
field lying either parallel or perpendicular to the plane of
incidence. Any diffraction effect or polarization dependence are thus
neglected.

The complex reflection and transmission coefficients (i.e., the ratios
of the reflected and the transmitted amplitudes to the incident one,
respectively) are given by~\cite{Yeh:2005eu}
\begin{equation}
  \label{eq:RTampFP}
  R (\Phi) = \frac{r[1-\exp(-i2\Phi)]}{1-r^2\exp(-i2\Phi)} \, ,  
  \quad
  T (\Phi) =\frac{(1-r^2)\exp(-i\Phi)}{1-r^2\exp(-i2\Phi)} \, .
\end{equation}
Here, $r$ is the Fresnel reflection coefficient for a wave travelling
from the surrounding medium into the FP and
\begin{equation}
  \Phi = \frac{2 \pi}{\lambda}  n d \cos \theta
\end{equation}
is the plate phase thickness, with $\lambda$ the wavelength in
vacuum and  $\theta$  the angle of refraction in the medium $n$, which
is related to the angle of incidence according to Snell's law.

The usual analysis proceeds by calculating the reflectivity and
transmissivity (i.e., the ratios of the reflected and the transmitted
intensities to the incident one).   The expressions are obtained directly from
equation~(\ref{eq:RTampFP}) and read
\begin{equation}
  \label{eq:RTintFP}
  \mathcal{R} = |R|^{2} = \frac{F \sin^{2} \Phi}
  {1 + F \sin^{2} \Phi } \, ,
  \qquad
  \mathcal{T}  = |T|^{2} = \frac{1}{1 + F \sin^{2} \Phi} \, ,
\end{equation} 
where the parameter $F$ is
\begin{equation}
  F = \frac{4 |r|^{2}} {(1- | r |^{2} )^2} \, .
\end{equation} 
Although $r$ is up to now a real number, we formally treat it as a
complex for reasons that will become apparent soon and so $| r |^{2} $
is the reflectivity of the plate
surfaces. Equations~(\ref{eq:RTintFP}) constitute the time honored
Airy formulas. Evidently, since there are no losses, the two patterns
are complementary, in the sense that
\begin{equation}
  \label{eq:ComRT}
  \mathcal{R} + \mathcal{T} = 1 \, . 
\end{equation}

\begin{figure}
   \centering{\includegraphics[height=6cm]{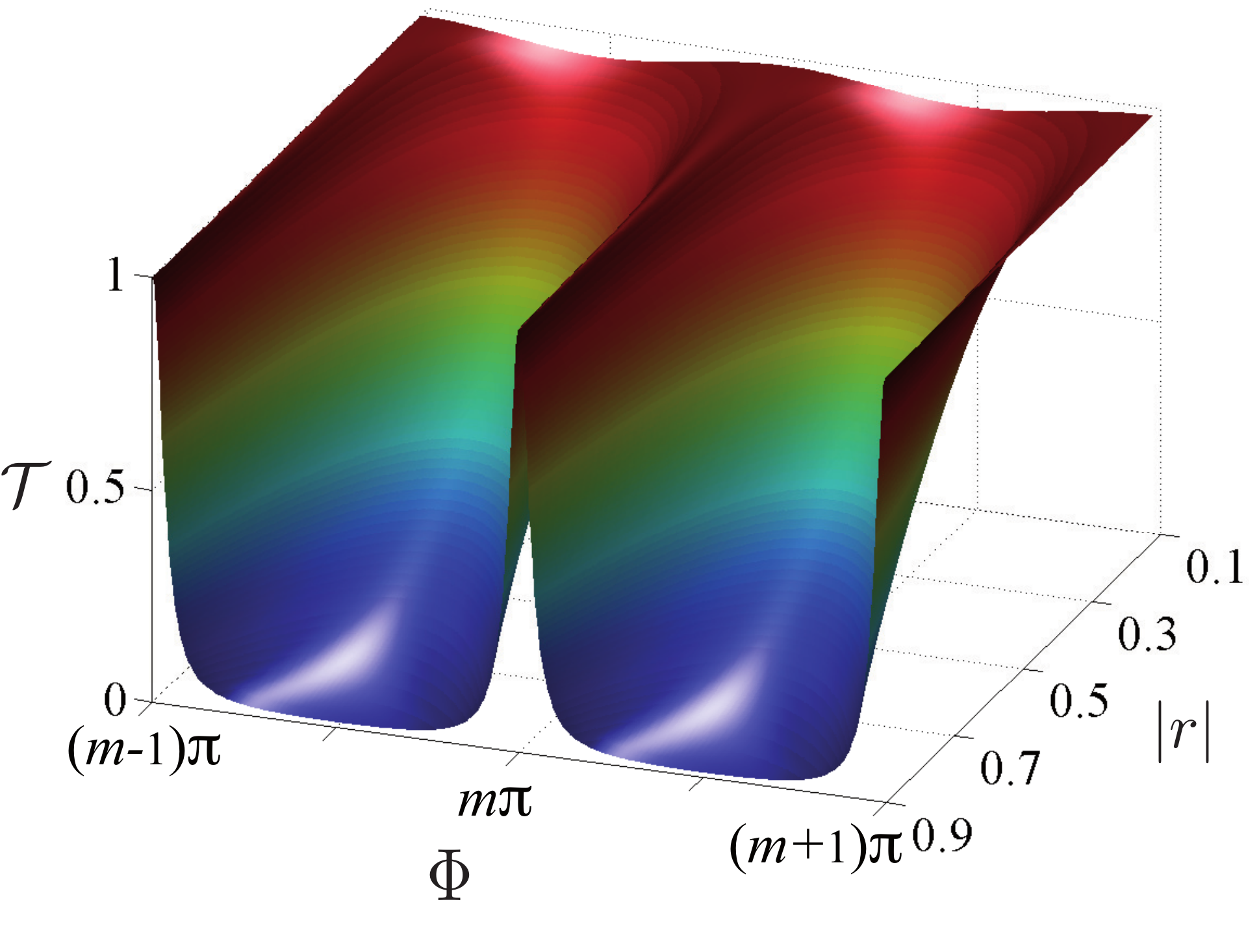}}
  \caption{Transmissivity $\mathcal{T}$ of the FP as a function of the
    phase shift $\Phi$ and the parameter $| r |$, whose square is the
    reflectivity of the plate surfaces.}
  \label{fig:TphiF}
\end{figure}

In figure~\ref{fig:TphiF} we plot the transmissivity $\mathcal{T}$ as a
function of the phase thickness $\Phi$ and $|r |$. As $| r |$
increases, the minima of $\mathcal{T}$ fall and the maxima become
sharper. In the limit of high $| r |$, the pattern consists on narrow
bright fringes on an almost completely dark background.

The sharpness of the fringes is conveniently measured by their full
width at half maximum (FWHM), which is the width between the points on
either side of a maximum where the intensity has fallen to half its
maximum value. The ratio of the separation of adjacent fringes (also
called free spectral range) and the FWHM is called the finesse
$\mathcal{F}$ of the fringes. A direct calculation shows that
\begin{equation}
  \mathcal{F} = \frac{\pi \sqrt{F}}{2} \, .
\end{equation}
This quantity is a measure of the apparatus ability to resolve closely
spaced spectral features. High values of $\mathcal{F}$ require an
increased reflectivity $|r|^{2}$ and this is accomplished by coating the
plates surfaces with a mirror. In what follows, we assume that such a
mirror is lossless.  In that case, the Airy formulas still hold
provided we interpret $r$ as the reflection coefficient of the mirror
(which becomes now a complex number). This adds to the plate phase
thickness $\Phi$ a phase change on the reflection at the mirrors.  In
general, both modulus and phase of the complex $r$ depend on the angle
of incidence and the dispersion properties of the material, albeit
such a variation can be disregarded for most practical purposes.

\section{Amplitude response of the Fabry-Perot}
 
Let us look in more detail at the amplitudes (\ref{eq:RTampFP}). First
of all, we observe that $R (\Phi)$ is a $\pi$-periodic function, while
$T(\Phi)$ is $2\pi$-periodic. Such a difference cannot be seen in the
intensity response, for both $\mathcal{R}$ and $\mathcal{T}$ have the
same period $\pi$.

To proceed further, let us rewrite equation~(\ref{eq:RTampFP}) as
\begin{eqnarray}
  \label{eq:quad}
  R  (\Phi) &=& \frac{2 | r | \sin\Phi}{N(\Phi)}  
  [ X ( \Phi) + i Y (\Phi) ] \, , \nonumber \\
  & & \\
  T  (\Phi) &=& \frac{1- | r |^{2}}{i N(\Phi)} 
  [ X (\Phi) - i Y (\Phi)  ] \, . \nonumber
\end{eqnarray}
where we have defined
\begin{eqnarray}
  & X (\Phi) = ( 1 + | r |^{2} ) \sin \Phi, 
  \qquad
  Y (\Phi) =  ( 1 - | r |^{2} )  \cos \Phi\, , & \nonumber \\
  & &  \\
  & N (\Phi) = (1- | r |^{2} )^2 + 4 | r |^{2} \sin^2 \Phi \, . &  
  \nonumber
\end{eqnarray}
We recall that any harmonic signal  $x(t)$ can be decomposed as
\begin{equation}
  \label{eq:comq}
  x(t) = X \, \cos \omega t + Y \, \sin \omega t \, ,
\end{equation}
where the $X$ and $Y$ are the in-phase and ($\pi/2$) out-of-phase
quadratures. In this spirit, $X(\Phi)$ and $Y(\Phi)$ can be seen as
sort of quadratures for $R(\Phi) $ and $T (\Phi)$. We stress, however,
that at difference of the harmonic oscillator, here the pre-factors in
their definition (\ref{eq:quad}) are not constant, but depend on
$\Phi$. This arises because the complex amplitudes $R (\Phi) $ and
$T (\Phi)$ do not have constant modulus, as in the oscillator.

\begin{figure}[b]
 \centering{\includegraphics[height=5.5cm]{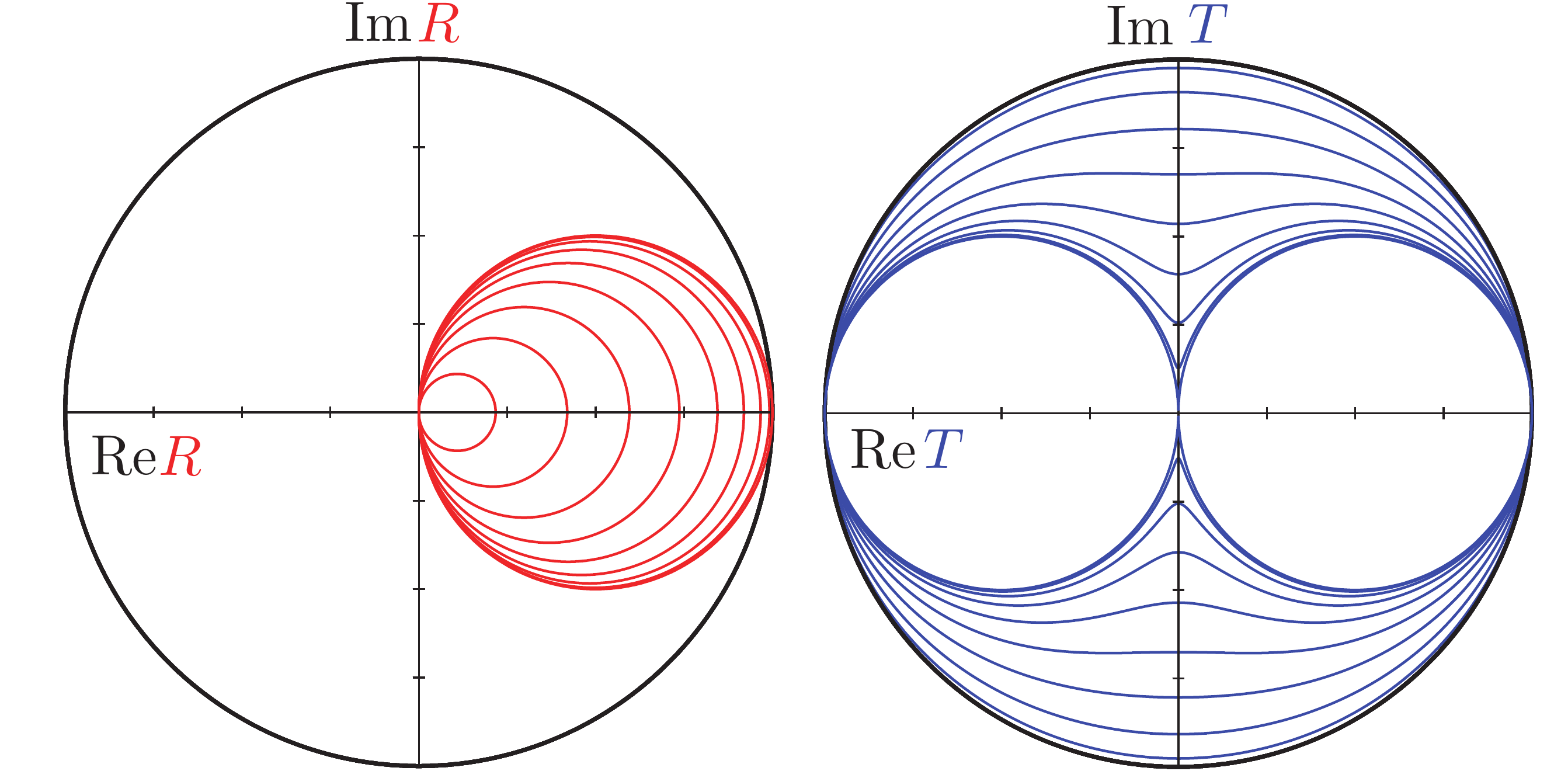}}
  \caption{Phase-space trajectories for $R(\Phi)$ (left) and $T(\Phi)$
    (right) as given in equation~(\ref{eq:quad}). They lie inside the unit
    disk.  The different curves correspond to different values of the
    plate reflectivity ranging from $r  = 0.11$ to $r = 0.99 $ in
    steps of 0.11 For $R(\Phi)$, the curves increase in size with $|r|$,
    while the converse happens for $T (\Phi)$.}
  \label{fig:RTphas}
\end{figure}

In figure~\ref{fig:RTphas} we represent these complex amplitudes for
several values of $|r|$. The different behavior commented above in
relation with the periodicity translates into the fact that when $T
(\Phi)$ completes a revolution, $R(\Phi) $ makes two turns.  Both
amplitudes lie inside the unit disk because of the fundamental
constraint~(\ref{eq:ComRT}).

To get a better grasp of these amplitudes, we introduce polar
coordinates as
\begin{equation}
  R = |R| \exp (i \rho) \, , 
  \qquad
  T = |T| \exp (i \tau) \, . 
\end{equation}
With this parametrization, we can recast equations~(\ref{eq:quad}) as
\begin{equation}
  \label{eq:param}
  \mathcal{R}  =  4 a^{2} \cos^{2} \rho  \, , 
  \qquad
  \mathcal{T}   =  1 -  4 a^{2}\sin^2 \tau  \, ,
\end{equation}
where we have used the definition in (\ref{eq:RTintFP}) and $a$
is a real parameter 
\begin{equation}
  \label{eq:defa}
  a = \frac{|r|}{1 + |r|^{2}} \, ,
\end{equation}
so that $ 0 \le a \le 1/2$.  

In this way, the reflected amplitude $R (\Phi) $ can be immediately
identified as a circle of radius $a$ centered in the point $(a, 0)$ of
the real axis.

The transmitted amplitude $T (\Phi) $ describes a hippopede (which
literally means ``horse fetter''). It was first investigated by
Proclus~\cite{Proclus:1992fq} and later on by
Booth~\cite{Booth:1877zt}, hence their names are sometimes attached to
this stunning curve.  For $0 < a < 1/\sqrt{8}$ it is an oval, and for
$1/\sqrt{8} < a < 1/2$ it is an indented oval, which tends to be an
eight in the limit $a=1/2$ (i.e., $|r | \rightarrow 1$). Yet not so
well known in Physics, it has a truly amazing set of properties that
the reader can look up in the abundant literature on the
subject~\cite{Ferreol:qr,Coffman:fc,wassenaar:cr}. We merely quote
that the hippopede can be defined as the curve formed by the
intersection of a torus and a plane parallel to the axis of the torus
and tangent to it on the interior circle. It is thus a spiric
section~\cite{Brieskorn:1986hs}.

For every value of $| r |$, the reflected amplitude passes through the
origin: $R(\Phi)$ is zero for $\Phi=0$ and $\pi$ and traces the circle
clockwise, getting its maximum  at $\Phi = \pm \pi/2$, where
$\rho = 0$ and then, according equation~(\ref{eq:param}),
$\mathcal{R}_{\mathrm{max}} = 4 a^{2}$.

On the other hand, the transmitted amplitude also describes the
hippopede clockwise. At $\Phi = 0$ and $\pi$, $T (\Phi)$
reaches its maxima, which are in the real axis at $T =1$ and $-1$,
respectively.  The minimum occurs at $\Phi = \pi/2$ and $3 \pi/2$,
where $\tau = -\pi/2$ and $-3 \pi/2$, respectively.  Therefore $
\mathcal{T}_{\mathrm{min}} = 1 - 4a^{2}$, which corresponds to half
the waist of the hipoppede at its indentation.

\begin{figure}[b]
  \centering{\includegraphics[height=5.5cm]{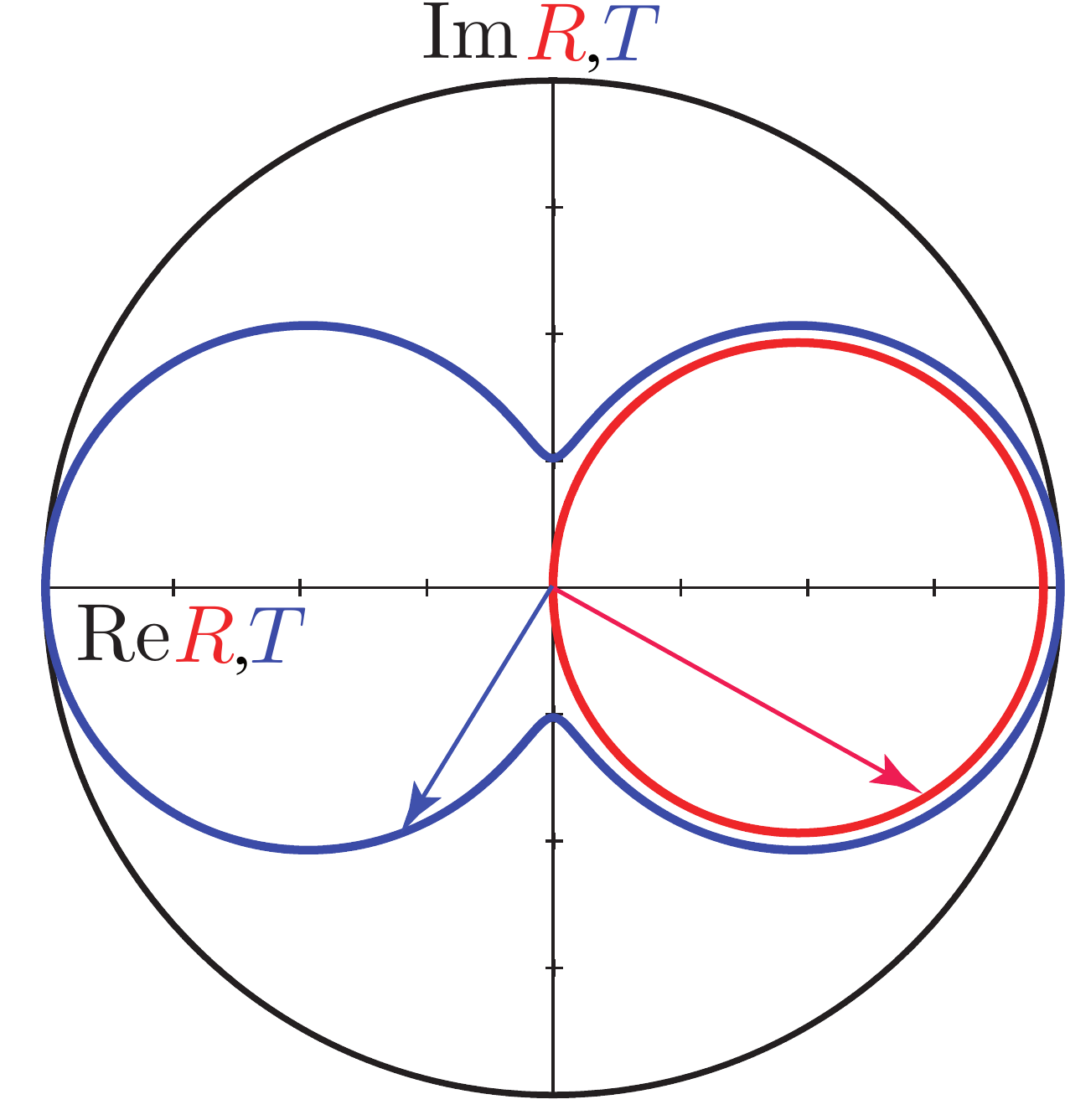}}
  \caption{Reflected (red) and transmitted (blue) amplitudes $ R(\Phi)
    $ and $ T (\Phi) $ for a transparent FP with $| r | = 0.77$. For
    every fixed value of the parameter $\Phi$ the corresponding
    position vectors are orthogonal.}
  \label{fig:Ortho}
\end{figure}

One can  check that
\begin{equation}
  \frac{R(\Phi)}{T (\Phi)}=   i \sqrt{F}  \sin \Phi \, ,
\end{equation}
which, in turn, implies that for a transparent symmetric system, as
the one we are dealing with, we have
\begin{equation}
  \label{eq:tau-rho}
  \rho (\Phi) - \tau (\Phi)  = \pm \frac{\pi}{2} \, .
\end{equation}
Consequently, for every value of $\Phi$ the reflected and transmitted
amplitudes are at quadrature. This is illustrated in
figure~\ref{fig:Ortho}, where we see that the position vectors of
$R(\Phi)$ and $T(\Phi)$ are orthogonal, as it is somehow implicit in
equation~(\ref{eq:quad}). Apropos of this, it is noteworthy to mention that
quite similar relations may be derived under the general assumptions
of symmetric and lossless
systems~\cite{Degiorgio:1980fk,Zeilinger:1981fv,Ou:1989pd}.

\begin{figure}
  \centering{\includegraphics[height=5.5cm]{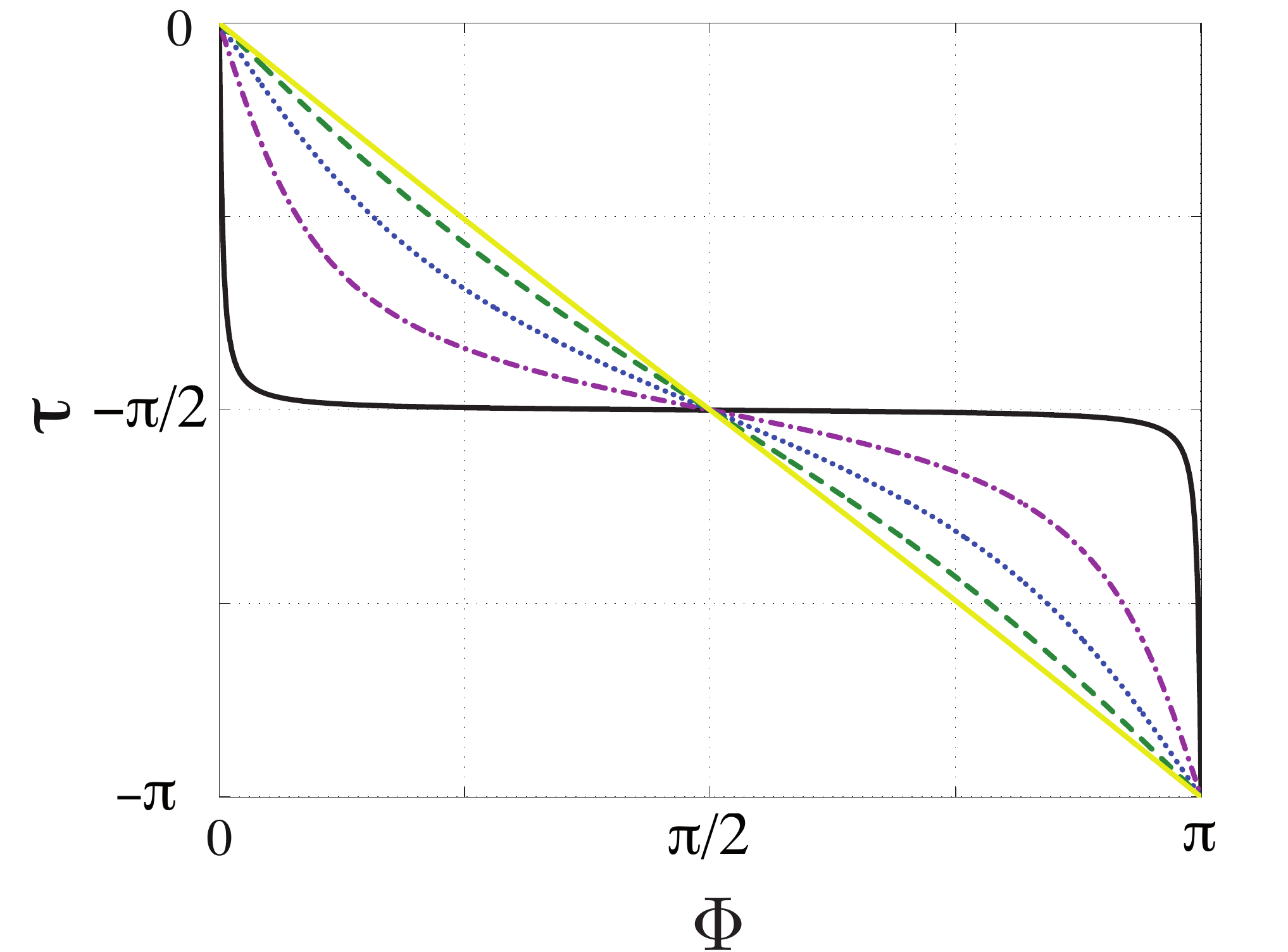}}
  \caption{Transmitted phase lag $\tau$ by an FP as a
    function of $\Phi$ for different values of $|r|$, from 
 0.11 (yellow full line)  to 0.99 (black full line) in steps of
 0.22. The curves bend more as $| r |$ increases.} 
  \label{fig:vel}
\end{figure}

We next examine the local slopes $\dot{\rho}(\Phi)$ and $\dot{\tau}
(\Phi)$, the dot denoting derivative respect to the parameter.  They
are the ``rates'' at which the curves are traced out. Indeed, they
entail a valuable physical interpretation. If we focus for simplicity
at $\tau$, we can write
\begin{equation}
  \label{eq:dpm}
  \frac{d\tau}{d \omega} = \dot{\tau} (\Phi) \,   
 \frac{d\Phi}{d\omega} \, , 
\end{equation}
with $\omega$ being the angular frequency. Now, $d\Phi/d\omega$ is the
single-pass time inside the cavity medium (for a non-dispersive
material, this is $nd\cos \theta/c$, where $c$ is the velocity of
light in vacuum), and $d \tau/d \omega$ is the time flight through the
FP, which incorporates the feedback. Hence, $ d\tau/d \omega$ may be
viewed as an enhancement factor of the time flight due to the
FP~\cite{Yariv:2006ab,Schwelb:2004fb}.

Because of equation~(\ref{eq:tau-rho}), both are equal for a lossless
medium: $ \dot{\rho} (\Phi) =\dot{\tau}(\Phi)$. In addition, we have
\begin{equation}
  \label{eq:2}
  \frac{\dot{\tau}}{\mathcal{T}} = 
  - \frac{1 + |r|^{2}}{1 - |r|^{2}}  \, ,
\end{equation}
where the negative sign indicates that the curve is oriented
clockwise.  This quotient is thus independent of $\Phi$: where the
transmitted amplitude is large, so is the velocity and the opposite.

To gain further insight into this issue, in figure~\ref{fig:vel} we plot
$\tau$ as a function of $\Phi$ for several values of $|r|$. Note that
the range of variation of $\tau$ is from $- 2 \pi$ to 0 when $\Phi$
goes from 0 to $2 \pi$,  but $\rho$
ranges from $-\pi/2$ to $\pi/2$, as one can directly infer at a glance
from figure~\ref{fig:Ortho}.  When $|r|$ is small, $\tau$ is almost a
straight line, with a slope pretty constant: this is the case when the
hippopede is almost an oval, without indentation. However, as $|r|$
increases, $\tau$ starts bending near  $\Phi = - \pi/2$, which is
precisely  at the waist. In the limit $|r|
\rightarrow 1$, $\tau$ becomes almost horizontal, with slope zero
almost everywhere, except a narrow interval around the maxima of
$\mathcal{T}$, where it quickly gets large.
 
\begin{figure}
  \centering{\includegraphics[height=5.5cm]{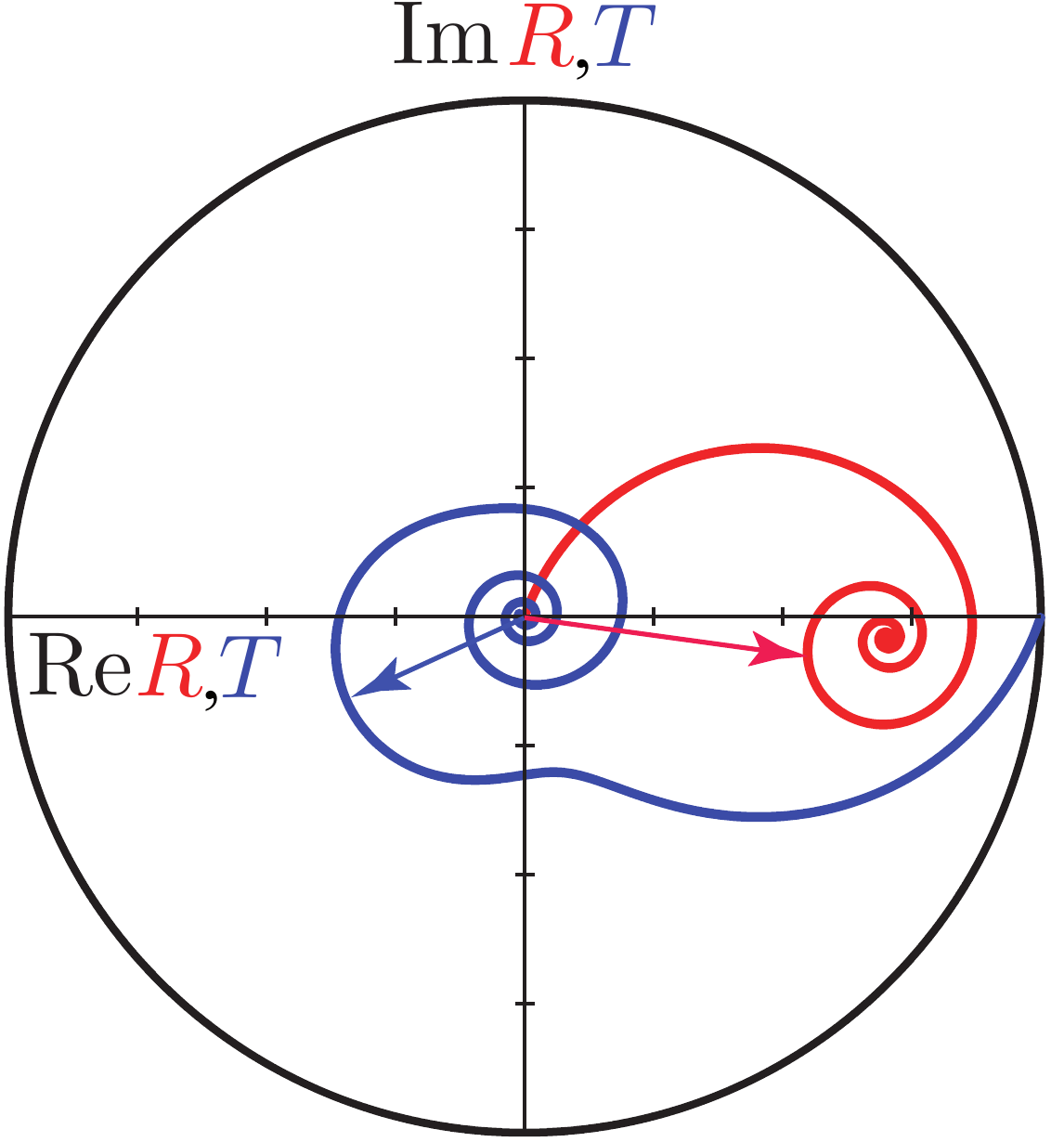}}
  \caption{Same as in figure~\ref{fig:Ortho}, but for an absorbing FP
    made of germanium with complex refractive index
    $N= 5.588 - i \ 0.933$ at a wavelength of 0.6199 $\mu$m.  We take
    normal incidence and the film thickness $d$ varying between 0 and
    0.35~$\mu$m. The marked position vectors correspond to
    $d= 0.052$~$\mu$m.}
  \label{fig:loxo}
\end{figure}
The previous considerations can be extended to a lossy (or gain)
cavity medium, a case of particular interest in laser physics.  The
medium  is now specified by a complex refractive index (and so a
complex $\Phi$), whose imaginary part accounts for the losses. The
resulting trajectories, for the simple case of a plate of germanium,
are shown in figure~\ref{fig:loxo} and they turn out to be
loxodromics~\cite{Monzon:2011kx}, a universal feature of absorption.
They start at $R=0$ and $T=1$ (when there is no film) and tend to $R
\rightarrow r$ and $T \rightarrow 0$ (when the film becomes
opaque). We can appreciate that the orthogonality between position
vectors does not hold true, neither the complementary relation
equation~(\ref{eq:ComRT}).

\section{Concluding remarks}

In summary, we have thoroughly explored the amplitude response of an
FP interferometer. Despite its basic nature and its simplicity, this
topic has been snubbed in the literature, which has strengthened
instead the role of the associated intensity.

Given the relevant role played by the Airy formulas, one might have
expected that their ``square root'' counterparts should have
remarkable features. This is indeed the case, as our results indicate:
the reflected amplitude traces a circle, while the transmitted one is
a hippopede, an intriguing curve full of nice mathematical properties.

We finally stress that  the phase-space methods employed here are
quite appealing for they have branched into offshoots of  importance
for modern physical theories. 

\bigskip

\ack

Many of the ideas in this paper originated from a long cooperation
with the late Alberto G. Barriuso, who unexpectedly passed away before
being able to guide this work to completion.  This paper is dedicated
to his memory.  Over the years, these ideas have been further
developed and expanded with questions, suggestions, criticism, and
advice from many colleagues. Particular thanks for help in various
ways goes to G. Bj\"{o}rk, J. F. Cari\~{n}ena, H. de Guise, P. de la
Hoz, A. B. Klimov, G. Leuchs, and J. M. Montesinos-Amilibia.  This
work is partially supported by the Spanish MINECO (Grant
FIS2011-26786).

\newpage


\end{document}